\documentstyle[mathrsfs,pra,aps]{revtex}

\begin{document}
\draft

\def\overlay#1#2{\setbox0=\hbox{#1}\setbox1=\hbox to \wd0{\hss #2\hss}#1%
\hskip
-2\wd0\copy1}
\twocolumn[
\hsize\textwidth\columnwidth\hsize\csname@twocolumnfalse\endcsname

\title{Quantum backaction of optical observations on Bose--Einstein
condensates}
\author{U.\ Leonhardt$^1$, T.\ Kiss$^2$, and P.\ Piwnicki$^3$}
\address{~$^1$Physics Department, Royal Institute of Technology (KTH),
Lindstedtsv\"agen 24, S-10044 Stockholm, Sweden}
\address{~$^2$Department of Nonlinear and Quantum Optics,
Institute for Solid State Physics and Optics,
P.O. Box 49, H-1525 Budapest, Hungary}
\address{~$^3$Abteilung f\"ur Quantenphysik, Universit\"at Ulm,
D--89069 Ulm, Germany}
\maketitle
\begin{abstract}
Impressive pictures of moving Bose--Einstein condensates have been taken
using phase--contrast imaging {[}M. R. Andrews {\it et al.},
Science {\bf 273}, 84 (1996){]}.
We calculate the quantum backaction of this measurement technique,
assuming the absence of residual absorption.
We find that the condensate gets gradually depleted at a universal
rate that is proportional to the light intensity and to
the inverse cube of the optical wave length.
The fewer atoms are condensed the higher is the required intensity
to see a picture, and, consequently, the higher is the induced
backaction.
To describe the quantum physics of phase--contrast imaging
we put forward a new approach to quantum--optical propagation.
We develop an effective field theory of paraxial optics
in a fully quantized atomic medium.
\end{abstract}
\date{today}
\maketitle
\vskip2pc]
\narrowtext
\section{Introduction}

Bose--Einstein condensates are macroscopic quantum states of many atoms with
nearly identical wave functions.
In addition to their prospects as laser--like sources of matter,
the condensates will offer a new fascinating testing ground for the
fundamentals of quantum mechanics, because of their macroscopic
yet quantum nature.
In all experiments performed so far
since the pioneering breakthrough \cite{pioneers}
Bose condensates are observed optically.
One particularly intriguing method is phase--contrast imaging
\cite{Andrews,Hulet}.
This technique allows to take impressive {\it in situ} pictures of moving
droplets of Bose--condensed matter, similar to phase--contrast images
of delicate living cells \cite{Hecht}.
Figure 1 shows a simplified scheme of the method.
Non--resonant laser light illuminates the sample, travels through,
and attains phase shifts that are proportional to the density of the
condensate. The remaining part of the apparatus serves to measure
the acquired phase gradient using the non--scattered part of the incident
light as a reference. Since the light is non--resonant,
the condensate is hardly disturbed and, indeed, many snapshots
or even entire video sequences of an individual sample have been taken
without significant effect \cite{Andrews}.

One might suspect \cite{Andrews}, however, that the observation will
nevertheless cause a quantum backaction on the condensate.
Imagine that phase--contrast imaging were an ideal von--Neumann
measurement of atomic positions.
After each measurement the atoms were frozen in position eigenstates
and their future motion were drastically altered.
Why is the observed backaction much less dramatic?
How does phase--contrast imaging affect the condensate?
The authors of the method \cite{Andrews} state that
``although dispersive scattering does not heat up the cloud
and destroy the condensate, it will change the phase as a
result of frequency shifts by the ac Stark effect.''
This would still allow ``a nonperturbative measurement of the
number of condensed atoms, which is the variable complementary
to the phase (so--called quantum nondemolition measurement).''

In this paper we find that 
two processes contribute to the backaction.
One is the phase diffusion mentioned above \cite{Andrews}
and the second is a depletion of the condensate.
The authors of Ref.\ \cite{Andrews}
have already seen indications of a gradual depletion.
However, they have attributed this to a residual light absorption.
We find it likely that their qualitative experimental
findings do in reality indicate the effect of the quantum backaction.

The physical reason for the quantum backaction is the local
interaction of the meter (the incident light) with the object (the
condensate). Therefore, to understand the quantum nature of
phase--contrast imaging, we must study the quantum propagation of light
in Bose--Einstein condensates. To a remarkable degree of accuracy
\cite{bec_optics} Bose condensates act simply as dielectric media on
the incident light. In the limit of a large detuning from optical
resonances of the atoms we can neglect the imaginary part of the
refraction index, i.e. we consider the condensate a lossless and
dispersionless dielectric medium. Quantum--optical propagation in bulk,
lossless media has been studied already in detail \cite{Media}.
A bulk medium, however, is hardly affected by light traveling
through, and the quantum backaction was rightfully ignored so far
\cite{Media}. In contrast to bulk matter,
a delicate quantum gas may feel the disturbances caused
by incident light. To understand {\em actio} and {\em reactio} of
light in quantum gases we essay a quantum theory of paraxial light
propagation in fully quantized, bosonic media.
We start from the safe ground of canonical electromagnetism in
dielectric media. Then we develop an approximate theory that allows an
analytic calculation of the backaction. Our theory is inspired by
the traditional paraxial approximation \cite{Marte} to study classical light
propagation. We quantize this model and conclude Section II with a
discussion of the paraxial field commutator, a critical quantity that
enters the backaction rates. Section II thus develops a general,
phenomenological theory of non--resonant light propagation in matter
waves.

Section III is more specific. Here we derive the master equation that
describes the backaction of the propagating light on the
condensate. The light plays a double role: It acts both as a meter and
as a reservoir. We apply the Born--Markov approximation of reservoir
theory \cite{Carmichael} to derive a general master equation that
describes the evolution of all (condensed and non--condensed)
atoms. Then we restrict our attention to the condensate and analyze
the two dissipative processes that turn out to appear due to the
quantum backaction: phase diffusion and atomic depletion.
We conclude Section III estimating the order of magnitude of the two
effects. We find that the depletion far outweighs the phase
diffusion. Furthermore, our estimation seems to indicate that the
depletion is experimentally significant.
Condensates with few atoms are especially vulnerable to
optical observations, in agreement with experimental evidence
\cite{Hulet}.

\section{Light and matter waves}

\subsection{Canonical theory}

Let us start from the canonical theory of electromagnetism in a
dielectric medium made of a single matter wave $\psi$. The total
Lagrangian density of light and matter is (in SI units)
\begin{equation}
  \label{eq:lg}
  {\mathscr L} = \frac{\epsilon_0}{2} (E^2-c^2 B^2) + \frac{\epsilon_0
  \chi_0}{2} E^2 |\psi|^2+{\mathscr L}_M \, .
\end{equation}
The first term describes a free electromagnetic field characterized by
the potentials $U$ and ${\mathbf A}$ that constitute the field
strengths
\begin{equation}
  {\mathbf E}=-\frac{\partial {\mathbf A}}{\partial t} - \nabla U \, ,
  \quad {\mathbf B}=\nabla\times {\mathbf A} \, .
\end{equation}
The second term plays a double role. The term describes both the
effect of the medium $|\psi|^2$ on the light $E^2$ and the backaction
of the light on the medium. The third term
\begin{equation}
    {\mathscr L}_M = \frac{i \hbar}{2}(\psi^* \dot{\psi}-\dot{\psi}^*
    \psi)- \frac{\hbar^2}{2 m} (\nabla \psi^*)(\nabla \psi) - V |\psi|^2
\end{equation}
characterizes the center--of--mass motion of the matter wave in the
external potential $V$. One can show that the Euler--Lagrange equations
generated by minimizing the action $\int {\mathscr L}\, d^4 x$ are indeed
Maxwell's equations of light in a dielectric medium with
susceptibility $\epsilon_0 \chi_0 |\psi|^2$.
Additionally, we obtain the Schr\"odinger equation of a matter wave
moving in both the external potential $V$ and in the optical potential
$-\epsilon_0 \chi_0 E^2/2$. This proves that the Lagrangian density
$\mathscr L$ describes the physics of a single matter wave that interacts
non--resonantly with a classical electromagnetic field.

Traditional quantum optics operates mostly with Hamiltonians. To find
the Hamiltonian density $\mathscr H$ of our model we follow the canonical
procedure. We calculate the functional derivatives
\begin{eqnarray}
  \frac{\delta {\mathscr L}}{\delta \dot{\mathbf A}} &=& - \epsilon_0
  \left( 1+ \chi_0 |\psi|^2 \right) {\mathbf E} \equiv - {\mathbf D}
  \nonumber \\
  \frac{\delta{\mathscr L}}{\delta \dot{\psi}} &=& \frac{i \hbar}{2}
  \psi^*\, , \quad \frac{\delta {\mathscr L}}{\delta \dot{\psi}^*}= -
  \frac{i \hbar}{2} \psi
\end{eqnarray}
and get
\begin{equation}
  {\mathscr H} \equiv - {\mathbf D} \dot{\mathbf A}+ \frac{i \hbar}{2}
  (\psi^* \dot{\psi}-\dot{\psi}^*\psi)-{\mathscr L} =
  {\mathscr H}_F+{\mathscr H}_M
\end{equation}
with the atomic Hamiltonian density
\begin{equation}
  {\mathscr H}_M = \frac{\hbar^2}{2 m} (\nabla \psi^*)(\nabla \psi)+V|\psi|^2
\end{equation}
and the electromagnetic part
\begin{equation}
  {\mathscr H}_F = \frac{\mathbf D E}{2} + {\mathbf D} \nabla U+
  \frac{\epsilon_0 c^2}{2} {\mathbf B}^2 \, .
\end{equation}
Observing the Maxwell equation $\nabla {\mathbf D}=0$ we ignore
${\mathbf D} \nabla U = \nabla ({\mathbf D}U)$ in ${\mathscr H}_F$ and
arrive at the electromagnetic Hamiltonian density
\begin{equation}
  \label{eq:h}
  {\mathscr H}_F = \frac{{\mathbf D}^2}{2 \epsilon_0 (1+\chi_0 |\psi|^2)}+
  \frac{\epsilon_0 c^2}{2} {\mathbf B}^2 \, .
\end{equation}
This expression is rather unpleasant, since the interaction between
light and matter appears in the denominator $(1+\chi_0|\psi|^2)$.
Of course, we could represent $(1+\chi_0|\psi|^2)^{-1}$ as the
geometric series $\sum_{\nu=0}^{\infty} (-\chi_0|\psi|^2)^\nu$,
provided that the series converges. The lowest--order theory
\begin{equation}
  \label{eq:hlow}
  {\mathscr H}_R=\frac{{\mathbf D}^2}{2 \epsilon_0}+ \frac{\epsilon_0
  c^2}{2} {\mathbf B}^2-\frac{\chi_0}{2 \epsilon_0} {\mathbf D}^2 |\psi|^2
\end{equation}
has a familiar appearance in quantum optics, because the Hamiltonian
density ${\mathscr H}_R$ contains the biquadratic interaction term
${\mathbf D}^2 |\psi|^2$. However, if we were to use the lowest--order
${\mathscr H}_R$ in Section III to calculate the quantum backaction in
phase--contrast imaging, we were to face an infinite result that clearly
contradicts experimental evidence \cite{Andrews}. Can we formulate
light propagation in matter waves in a different way?

\subsection{Optical Schr\"odinger equation}

Our canonical field theory of light and matter contains Maxwell's
equations in the form of the Lagrangian $\mathscr L$ or, equivalently, in
the Hamiltonian $\mathscr H$.
In classical optics, however, one hardly uses Maxwell's equation to
describe paraxial light propagation, but applies a characteristic
approximation which is an optical analogue of the Schr\"odinger equation.
Let us sketch a
brief derivation of the optical Schr\"odinger equation \cite{Marte}.
We start from the wave equation
\begin{eqnarray}
&&\nabla^2{\mathbf E}- \frac{1+\chi_0|\psi|^2}{c^2} \frac{\partial^2 {\mathbf
E}}{ \partial t^2}=\nabla(\nabla{\mathbf E})\, ,\nonumber \\
&&\nabla {\mathbf E}=- \nabla \left[ \ln (1+\chi_0|\psi|^2)\right] {\mathbf
E}
\end{eqnarray}
(a familiar consequence of Maxwell's equations). We assume the medium $\chi_0
|\psi|^2$ to be weak and to be gradually varying in space. (A mirror, for
example, is excluded from this model.)
We neglect the polarization mixing $\nabla(\nabla {\mathbf E})$ and consider
paraxial propagation in the $z$ direction with carrier frequency $\omega_0$.
In the following we will often refer to the wave number $k_0=\omega_0/c$ and
to the wave length $\lambda = 2 \pi/k_0$.
We use the ansatz
\begin{equation}
{\mathbf E}^{(+)} = \mbox{\boldmath{$\cal E$}} \exp (i k_0 z -i \omega_0 t)
\end{equation}
for the positive frequency component ${\mathbf E}^{(+)}$ of the electric
field, and differentiate
\begin{eqnarray}
\frac{\partial^2 {\mathbf E}^{(+)}}{\partial t^2}
&\approx& \exp (i k_0 z - i \omega_0 t)\left(-2 i
\omega_0 \frac{\partial}{\partial t}-\omega_0^2\right)
\mbox{\boldmath{$\cal E$}}
\nonumber \\
\frac{\partial^2 {\mathbf E}^{(+)}}{\partial z^2}
&\approx& \exp (i k_0 z - i \omega_0 t)\left(+2 i
k_0 \frac{\partial}{\partial z}-k_0^2\right)
\mbox{\boldmath{$\cal E$}}
\end{eqnarray}
neglecting the slow variation of the envelope $\mbox{\boldmath{$\cal E$}}$.
Furthermore, we regard
\begin{equation}
\chi_0 |\psi|^2 \frac{\partial^2 {\mathbf E}}{\partial t^2} \approx -
\omega_0^2 \chi_0 |\psi|^2 {\mathbf E}
\end{equation}
and obtain
\begin{equation}
\left[ \nabla^2_\perp+2 i k_0 \left( \frac{\partial}{\partial
z}+\frac{1}{c} \frac{\partial}{\partial t} \right) +k_0^2 \chi_0 |\psi|^2
\right] \mbox{\boldmath{$\cal E$}} = 0
\end{equation}
with $\nabla^2_\perp$ being the transversal Laplacian
$\partial^2/\partial x^2+\partial^2/ \partial y^2$.
Finally, we utilize that $(\partial/\partial z +\partial / \partial ct)\exp
(i k_0 z- i \omega_0 t)$ vanishes and arrive at the optical Schr\"odinger
equation
\begin{equation}
\label{eq:os}
i \left(\frac{\partial}{\partial z}+ \frac{1}{c} \frac{\partial}{\partial t}
\right) {\mathbf E}^{(+)}= - \frac{\nabla^2_\perp}{2 k_0} {\mathbf
E}^{(+)} - \frac{k_0 \chi_0 |\psi|^2}{2} {\mathbf E}^{(+)} \, .
\end{equation}
More details about the optical Schr\"odinger equation (and beyond)
are elaborated in the comprehensive paper \cite{Marte}
by Marte and Stenholm.

\subsection{Canonical theory of paraxial propagation}

Let us describe the paraxial propagation of light in a canonical field
theory. We consider a fixed polarization, i.e. a scalar electrical field
strength $E^{(+)}$, and express $E^{(+)}$ as
\begin{equation}
E^{(+)}({\mathbf x},t)= \left(\frac{\hbar \omega_0}{2
\epsilon_0}\right)^{1/2} i \varphi({\mathbf x},t) \, .
\end{equation}
The complex function $\varphi$ represents the optical field in units of the
vacuum noise $(\hbar \omega_0/2 \epsilon_0)^{1/2}$. One can easily verify
that the Lagrangian density
\begin{eqnarray}
{\mathscr L}_P &=& {\mathscr L}_F + {\mathscr L}_M \nonumber \\
{\mathscr L}_F &=& \frac{i \hbar c}{2} \varphi^* \left(\frac{\partial}{\partial
z} +\frac{1}{c} \frac{\partial}{\partial t} \right) \varphi-
\frac{i \hbar c}{2} \varphi \left( \frac{\partial}{\partial z}+ \frac{1}{c}
\frac{ \partial}{\partial t} \right) \varphi^* \nonumber \\
&&- \frac{\hbar c}{2 k_0}
(\nabla_\perp \varphi^*)(\nabla_\perp \varphi)+\frac{\hbar
\omega_0}{2} \chi_0|\varphi|^2 |\psi|^2
\end{eqnarray}
generates the optical Schr\"odinger equation (\ref{eq:os}) as an
Euler--Lagrange equation for $\varphi$.
This proves that ${\mathscr L}_F$ is a valid Lagrangian to describe paraxial
optical propagation.
Lagrangians are defined up to a positive prefactor (because we are only
interested in their minimum). We have chosen the prefactor of ${\mathscr L}_F$
such that the coupling term $\hbar \omega_0 \chi_0
|\psi|^2|\varphi|^2/2=\epsilon_0\chi_0 |E|^2 |\psi|^2 \approx \epsilon_0
\chi_0 (E^{(+)}+E^{(-)})^2|\psi|^2/2$ is identical to the interaction term
$\epsilon_0 \chi_0 E^2 |\psi|^2/2$ in the canonical Lagrangian density
(\ref{eq:lg}). This is necessary, because the term plays the double role of
describing {\em actio} and {\em reactio} of light and matter.

In order to find the Hamiltonian density we follow the royal road of
canonical field theory. We calculate the functional derivatives
\begin{equation}
\frac{\delta {\mathscr L}}{\delta \dot{\varphi}} =\frac{i \hbar}{2}
\varphi^* \, , \quad
\frac{\delta {\mathscr L}}{\delta \dot{\varphi}^*}=-\frac{i \hbar}{2}\varphi
\end{equation}
and obtain
\begin{eqnarray}
\label{eq:hd}
{\mathscr H}_P &=& \frac{i \hbar}{2} \left( \varphi^*
\dot{\varphi}-\dot{\varphi}^* \varphi+\psi^*\dot{\psi}-\dot{\psi}^* \psi
\right) -{\mathscr L}_P \nonumber\\
&=& {\mathscr H}_M+{\mathscr H}_F \, ,\nonumber \\
{\mathscr H}_F &=& \frac{i \hbar c}{2}
\left( \frac{\partial \varphi^*}{\partial
z} \varphi - \varphi^* \frac{\partial \varphi}{\partial z} \right) +
\frac{\hbar c}{2 k_0}
(\nabla_\perp \varphi^*)(\nabla_\perp \varphi)\nonumber \\
&& -\frac{\hbar \omega_0}{2} \chi_0|\varphi|^2 |\psi|^2 \, .
\end{eqnarray}
The last term, $-\hbar \omega_0\chi_0|\varphi|^2 |\psi|^2/2$, has the
familiar appearance of a quantum--optical light--matter interaction. The other
terms of ${\mathscr H}_F$ describe the free paraxial propagation of
the optical field.

\subsection{Quantum statistics}

So far we have formulated the canonical theory of a single matter wave that
interacts non--resonantly with an optical field.
Let us turn to the description of a Bose gas of many atoms interacting with
photons. The many--body problem is conveniently formulated in the language of
Second Quantization where operators $\hat\psi$ and $\hat \varphi$ describe
the matter and light field, respectively. The atoms are bosons
\begin{equation}
\left[ \hat\psi({\mathbf x}_1),\hat\psi^\dagger({\mathbf x}_2) \right] =
\delta^{(3)} ({\mathbf x}_1-{\mathbf x}_2) \, .
\end{equation}
To find the Hamiltonian $\hat H$ of our model we integrate the Hamiltonian
density ${\mathscr H}_P$ over three--dimensional space and apply partial
integration. We obtain
\begin{eqnarray}
\hat H &=& \hat H_R+ \hat H_I + \hat H_M \\
\label{eq:hr}
\hat H_R &=& \hbar c \int \hat \varphi^\dagger \left(-i
\frac{\partial}{\partial z }- \frac{\nabla^2_\perp } {2 k_0} \right) \hat
\varphi\, d^3 x\\
\hat H_I &=& - \frac{\hbar \omega_0}{2} \chi_0 \int \hat \psi^\dagger
\hat \psi \, \hat \varphi^\dagger \hat \varphi \, d^3 x \, .
\label{eq:hi}
\end{eqnarray}
The Hamiltonian $\hat H_R$ describes free paraxial light propagation, $\hat
H_I$ accounts for the interaction of light and matter and $\hat H_M$
describes the quantum gas of the atoms.
We include atomic collisions in our model by adding an atom--atom
interaction term to the potential $V$, and obtain for example
\cite{DGPS}
\begin{equation}
\hat H_M  =  \int \left( 
-\frac{\hbar^2}{2m}\,\hat \psi^\dagger \nabla^2 \hat \psi +
V \hat \psi^\dagger \hat \psi +
\frac{g}{2}\,\hat \psi^{\dagger 2}\hat \psi^2
 \right) d^3 x
\, .
\end{equation}
Note that our results turn out to be independent 
on the explicit form of $\hat H_M$, as long as $\hat H_M$ is able
to maintain a Bose--Einstein condensate.

An important issue in our theory is the optical field commutator
\begin{equation}
\label{eq:c}
C({\mathbf x}_1-{\mathbf x}_2) \equiv \left[ \hat \varphi({\mathbf x}_1), \hat
\varphi ^\dagger({\mathbf x}_2) \right] \, .
\end{equation}
If $C({\mathbf x})$ is a three--dimensional delta function we obtain
the optical Schr\"odinger equation (\ref{eq:os}) for $\hat\varphi$
from the Heisenberg equation
\begin{equation}
\label{eq:heisen}
\frac{\partial \hat \varphi}{\partial t} =
- \frac{i}{\hbar} \, [\hat\varphi , \hat H ] \,.
\end{equation}
Is the commutator a delta function?
The field operator $\hat \varphi$ is one polarization component of the
photon absorption operator $\hat{\mathbf V}({\mathbf x},t)$ of Mandel and
Wolf \cite{MWV}.
The commutator $C({\mathbf x})$ is then given in terms of the density
$d\mu({\mathbf k})$ of the employed optical modes as \cite{MWC}
\begin{equation}
\label{eq:cm}
C({\mathbf x}) = \frac{1}{(2 \pi)^3} \int e^{i {\mathbf k x}} d\mu({\mathbf
k}) \, .
\end{equation}
The crucial point is that the mode density $d\mu({\mathbf k})$ is restricted,
since we consider only those modes that obey paraxial propagation.
Therefore, we cannot regard $C({\mathbf x})$ as a three--dimensional delta
function, in general. However, the commutator behaves like
$\delta^{(3)}({\mathbf x}_1-{\mathbf x}_2)$ on paraxial test functions that
are supported on the spectrum of $C$
(that share the same region in ${\mathbf k}$ space).
To see this we introduce the Fourier transformations
$\tilde C({\mathbf k})$ and $\tilde f({\mathbf k})$ of the
commutator and a test function $f({\mathbf x})$, and calculate
\begin{eqnarray}
\int C({\mathbf x}_1-{\mathbf x}_2) &f({\mathbf x}_2)& \, d^3x_2\nonumber \\
&=& \frac{1}{(2 \pi)^3} \int \tilde C({\mathbf k}) \tilde f ({\mathbf k})
e^{i{\mathbf k x}_1} d^3 k \nonumber \\
&=&\frac{1}{(2 \pi)^3} \int \tilde f({\mathbf k}) e^{i {\mathbf k x}_1}
d^3 k\nonumber \\
&=&f({\mathbf x}_1) \, .
\end{eqnarray}
This is the defining property of a three--dimensional delta function.
In particular, the field operator $\hat \varphi$ itself is a paraxial test
function, and thus we can safely regard $C({\mathbf x})$ as
$\delta^{(3)}({\mathbf x}_1-{\mathbf x}_2)$ in integrals containing $\hat
\varphi$.
In this way we obtain indeed from the Heisenberg equation (\ref{eq:heisen})
the optical Schr\"odinger equation for $\hat \varphi$, as is easily verified.
This shows that our quantum theory of paraxial light propagation is
consistent with the classical theory.

On the other hand,
for test functions with a broad spatial spectrum we put
\begin{eqnarray}
d\mu({\mathbf k})&=& \delta (\omega({\mathbf k})/\omega_0-1)\,d^3 k
\nonumber \\
&=& \delta \left( k_z/k_0 +
{\textstyle \frac{1}{2}} (k_x^2+k_y^2) / k_0^2-1
\right)\,
 d^3 k
\, .
\end{eqnarray}
We obtain, for example, the commutator {\em per se}
\begin{eqnarray}
\label{eq:cb}
C({\mathbf x}) &=& \frac{1}{(2 \pi)^3} \int e^{i {\mathbf k x}} \cdot 1 \cdot
d \mu ( {\mathbf k})\nonumber \\
&=& \frac{k_0^2}{i(2 \pi)^2(z-i 0)}
\exp \left[\frac{i k_0}{2 z}(x^2 + y^2)+ik_0 z \right] \, ,
\end{eqnarray}
see Appendix A. The commutator resembles the propagator of a free--particle
wave, because $C$ satisfies
\begin{eqnarray}
\left( -i \frac{\partial}{\partial z} - \frac{\nabla^2_\perp}{2 k_0} \right)
C &=& k_0 C \nonumber \\
C(x,y,z \to 0) &=& \lambda^{-1}\, \delta(x)\,\delta(y) \, .
\end{eqnarray}
Since commutators are Green's functions, we would expect this property
from a paraxial quantum optics.

We have thus seen that the commutator acts
differently on paraxial and on broad--spectrum test functions. The two faces
of the commutator will play two different roles in the quantum backaction of
phase--contrast imaging.

\section{Quantum backaction}

\subsection{Phase--contrast imaging}

Phase--contrast imaging is a method to see transparent objects (living cells,
for example) that are otherwise invisible. A transparent body does not absorb
incident light but imprints a phase shift that is proportional to the local
density of the object. Manipulations in the focal plane of the observing lens
system transform the image of a phase object into an equivalent image of an
amplitude object \cite{Hecht}, see Fig. 1. In the absence of absorption the
light will not significantly affect a macroscopic body, but it might affect a
delicate quantum gas.

Formulated in the language of quantum measurement theory \cite{QMT}, the
light is a meter of the atomic object: The incident light interacts with the
atoms, becomes entangled with the atomic sample, and is finally measured in a
classical apparatus (lens system and detectors). The light field is also a
reservoir: The incident beam, say a plane wave, is scattered on the object
into many emerging modes that form an image.
The degree of collecting the scattered modes and extracting the imprinted
phase information determines the performance of the measurement. The overall
performance is ultimately limited by the quantum--optical phase fluctuations
of the incident light and by the performance of the apparatus.

On the other hand, the average quantum backaction (averaged over all
experimental runs) is solely determined by the entanglement between object
and meter. This means that the average backaction is entirely independent on
the performance of the measurement. We can average over the meter to
calculate the backaction on the object. This averaging leads to a dissipative
dynamics of the object, formulated as a quantum master equation
\cite{kinetics}.

\subsection{Master equation}

We describe the state of all atoms by the density matrix $\hat \rho_I$ in the
interaction picture
\begin{equation}
\hat \rho_I(t) = e^{i\hat H_Mt/\hbar} \hat \rho(t) \, e^{-i \hat H_Mt/\hbar}
\end{equation}
and employ standard reservoir theory \cite{Carmichael} to derive an equation
of motion for $\hat \rho_I$.
The interaction between light and matter is weak and, consequently, we can
apply second--order perturbation theory (the Born approximation
\cite{Carmichael}).
The incident light travels through the object in a time $\tau_c$ that is
simply given by the extension of the condensate divided by the speed of
light. After $\tau_c$ we do not expect the light to interact further with the
atomic sample. Therefore, the effective memory of the light--matter
interaction is roughly given by $\tau_c$ and is very short.
So, in addition to the Born approximation, we can apply the short--memory
(Markov) approximation using an integration time $\tau_0$ that is
much larger than the time of flight $\tau_c$ but is still smaller
than a characteristic time of the atomic sample.
We will see later in Sections C and D that our results do not depend on the
precise value of the memory time $\tau_0$, provided that $\tau_0$
exceeds $\tau_c$. This justifies {\it a posteriori} our
Markovian theory. We obtain in the Born--Markov approximation
\cite{Carmichael}
\begin{eqnarray}
\label{eq:master}
\frac{d\hat \rho_I}{dt} &=&
\frac{i}{\hbar}\, \mbox{tr}_R
[\hat \rho_I(t) \hat \rho_{IR}(t) , \hat H_{II}(t)] -
{\cal L} \hat \rho_I \,\,,
\nonumber \\
{\cal L} \hat \rho_I &=&
\frac{1}{\hbar^2} \int \limits_{t-\tau_0}^t \mbox{tr}_R
[ \hat H_{II}(t), [\hat H_{II}(t^\prime), \hat \rho_I(t)
\hat \rho_{IR}(t)]]\, dt^\prime
\end{eqnarray}
with
\begin{eqnarray}
\hat H_{II} &\equiv& - \frac{\hbar \omega_0}{2} \chi_0 \int \hat
\psi_I^\dagger \hat \psi_I \, \hat \varphi_I^\dagger \hat \varphi_I\,
d^3 x \, ,
\nonumber \\
\hat\psi_I({\mathbf x},t)
&\equiv& e^{-i \hat H_M t/\hbar} \hat \psi ({\mathbf x})
e^{i\hat H_M t / \hbar} \, , \nonumber \\
\hat\varphi_I({\mathbf x},t)
&\equiv& e^{-i \hat H _R t / \hbar } \hat \varphi ({\mathbf x})
e^{i \hat H_R t/ \hbar} \, ,
\end{eqnarray}
where $\hat \psi$ and $\hat \varphi$ are Schr\"odinger--picture operators,
and $\hat \rho_{IR}$ denotes the density matrix of the light field in the
interaction picture
\begin{equation}
\hat \rho_{IR}(t) =
e^{i \hat H_R t/\hbar} \hat \rho_R(t) e^{-i\hat H_R t/\hbar} \,.
\end{equation}
The incident light is a plane wave
\begin{eqnarray}
\label{eq:pw}
\mbox{tr}_R
\left\{
\hat \varphi^\dagger ({\mathbf x}_1) \hat \varphi ({\mathbf x}_2)
\hat \rho_R\right\}
= \frac{I}{\hbar \omega_0 c}\, e^{i k_0 (z_2-z_1)}&&
 \, ,  \nonumber\\
\mbox{tr}_R
\left\{
\hat \varphi^\dagger ({\mathbf x}_1) \hat \varphi^\dagger
({\mathbf x}_2) \hat \varphi ({\mathbf x}_1)
\hat \varphi ({\mathbf x}_2) \hat \rho_R
\right\}
= \mbox{const}&&
\, ,
\end{eqnarray}
that travels in the $z$ direction and
carries an intensity (energy flux) of $I$.
Apart from the value of the intensity we do not need to specify the quantum
state of the illuminating plane wave. The atomic sample is a thin phase
object, thin enough to neglect the optical diffraction inside (the
$\nabla^2_\perp /2k_0$ term in Eq.\ (\ref{eq:hr})
of the Hamiltonian $\hat H_R$).
This means that $\hat \varphi_I$ obeys the equation
\begin{equation}
i\left(\frac{\partial}{\partial z}+\frac{1}{c} \frac{\partial}{\partial t}
\right) \hat \varphi_I = 0
\end{equation}
with the obvious solution
\begin{equation}
\hat \varphi_I({\mathbf x}, t+t_0) =
\hat \varphi_I ({\mathbf x}-c t\, {\mathbf e}_z, t_0) \, .
\label{eq:nodiff}
\end{equation}
The total number of atoms is fixed, i.e.
\begin{equation}
\label{eq:ac}
[\,\hat \rho, \int \hat \psi^\dagger \hat \psi\, d^3 x \,] = 0 \, .
\end{equation}
We assume the existence of a Bose--Einstein condensate with wave function
$\psi_0 ({\mathbf x},t)$. To describe this phenomenon, we expand the atomic
annihilation operator $\hat \psi_I({\mathbf x}, t)$ in the interaction
picture into a complete, orthonormal set of atomic modes
\begin{eqnarray}
\label{eq:atoms}
\hat \psi_I({\mathbf x},t) &=& \sum_\nu \psi_\nu({\mathbf x},t) \hat a_\nu \,
, \nonumber\\
\int \psi_\mu^*({\mathbf x},t) \psi_\nu({\mathbf x},t) d^3 x &=& \delta_{\mu
\nu} \, , \nonumber\\
\sum_\nu \psi_\nu^*({\mathbf x}_1, t) \psi_\nu({\mathbf x}_2,t) &=&
\delta^{(3)}({\mathbf x}_1-{\mathbf x}_2) \, .
\end{eqnarray}
We assume that only the $\psi_0$ mode (the condensate) is significantly
populated, $\hat a_\nu \hat \rho_I=0$ for $ \nu \neq 0$.
This means that we consider the case of zero temperature and that we 
neglect the fluctuations \cite{fluctuations} of the
above--condensate part, for simplicity. 
For repulsive atom--atom interactions 
the spatial wave function $\psi_0$ of the condensate
is dominated by the balance between the interatomic repulsion 
and the expernal potential \cite{DGPS}.
The state of the condensate is
described by the reduced density matrix
$\hat \rho_0 \equiv \mbox{tr}_{AC} \hat \rho_I$.

For finding the master equation of the condensate state $\hat \rho_0$, we
average the master equation (\ref{eq:master}) with respect
to the above--condensate part. This procedure requires some lengthy yet
straightforward calculations that we prefer to present in Appendix B.
We obtain the result
\begin{eqnarray}
\label{eq:master0}
\frac{d \hat \rho_0}{d t} &=& -\mbox{tr}_{AC}{\cal L}\hat \rho_I =
-{\cal L}_1 \hat \rho_0 - {\cal L}_2 \hat \rho_0 \, , \\
{\cal L}_1 \hat \rho_0 &=& \Gamma_P \left( \hat a^\dagger_0 \hat a_0\, \hat
a^\dagger_0 \hat a_0\, \hat \rho_0 - \hat a^\dagger_0 \hat a_0\,
\hat \rho_0\, \hat a^\dagger_0 \hat a_0 \right) + \mbox{H.c.} \, , \\
{\cal L}_2 \hat \rho_0 &=& (\Gamma_L-\Gamma_P)
\left(
a^\dagger_0  \hat a_0 \,\hat \rho_0 -
\hat a_0\, \hat \rho_0\, \hat a^\dagger_0
\right) + \mbox{H.c.} \, , \\
\label{eq:gp}
\Gamma_P &=& \int \limits_{t-\tau_0}^t \int \int G\left[ {\mathbf x}-{\mathbf
x}^\prime-c (t-t^\prime){\mathbf e}_z \right] \nonumber\\
&&\times\,
\left| \psi_0 ({\mathbf x},t) \right|^2 \left| \psi_0 ({\mathbf x}^\prime,
t^\prime) \right|^2 \, d^3 x \, d^3 x^\prime \, d t^\prime \, , \\
\label{eq:gl}
\Gamma_L &=& \int \limits_0^{\tau_0} G(c \tau {\mathbf e}_z)\, d\tau \, ,\\
\label{eq:g}
G({\mathbf x}) &\equiv& \frac{\omega_0}{4} \frac{\chi_0^2}{\hbar c} I
C({\mathbf x}) e^{-i k_0 z} \, .
\end{eqnarray}
We see that no macroscopic light force acts on the condensate.
This is easy to understand, because light forces are caused by intensity
gradients, and in our case the incident light is a plane wave with uniform
intensity $I$.

We notice that two dissipative processes, ${\cal L}_1 \hat \rho_0$ and ${\cal
L}_2 \hat \rho_0$, contribute to the backaction. The first one is a phase
diffusion (see Section C) and the second one is a bosonic depletion (see
Section D). What is the physical origin of the two processes?

For forming an image of the condensate, the incident light is scattered on
the atoms. Therefore the light transfers momentum (and energy) to the atoms.
After the scattering a condensate atom may find itself in the condensate
again or it may be expelled, with the (complex) rates $\Gamma_P$ and
$\Gamma_L - \Gamma_P$, respectively. The atom that rejoined the condensate
will suffer from a loss in phase coherence, causing phase diffusion, and the
expelled atoms will gradually deplete the condensate. Let us look into the
two processes in more detail.

\subsection{Phase diffusion}

Let us study the phase--diffusion ${\cal L}_1 \hat \rho_0$ in the atomic
number--state basis of the condensate
\begin{eqnarray}
\langle m|{\cal L}_1 \hat \rho_0 |n \rangle
&=&
\left[i \mbox{Im} \Gamma_P \, (m^2-n^2)+
\mbox{Re} \Gamma_P \, (m-n)^2 \right]
\nonumber\\
&&\times\, \langle m|\hat \rho_0|n \rangle \, .
\end{eqnarray}
We see that the imaginary part of $\Gamma_P$ governs a Hamiltonian process,
i.e. a commutator of $\hat \rho_0$ with a Hermitian effective Hamiltonian
that is proportional to the atom number. We may call it a Kerr--type process.

Let us turn to the complementary picture where we
describe ${\cal L}_1 \hat \rho_0$ in the basis of phase states
\cite{Phase} $|\phi\rangle \equiv (2 \pi)^{-1} \sum_{n=0}^\infty
\exp(in\phi)\,|n\rangle$. We obtain
\begin{eqnarray}
\langle \phi_1| {\cal L}_1 \hat \rho_0 |\phi_2\rangle
&=&
\left[ - \mbox{Re} \Gamma_P \left( \frac{\partial}{\partial \phi_1} +
\frac{\partial}{\partial \phi_2} \right)^2 \right.  \nonumber\\
&&\left. -i \mbox{Im} \Gamma_P \left(
\frac{ \partial^2}{\partial \phi_1^2}-\frac{\partial^2}{\partial \phi_2^2}
\right) \right] \langle \phi_1|\hat \rho_0 |\phi_2 \rangle \, .
\end{eqnarray}
The first term is clearly a diffusion process of the atomic phase (and hence
justifies our terminology of calling ${\cal L}_1\hat\rho_0$ a phase
diffusion). The diffusion rate $\gamma_P$ is the real part of $\Gamma_P$.

To calculate $\gamma_P$ we utilize that the characteristic time of the
condensate is much larger than the time of flight $\tau_c$ of the incident
light and the integration time $\tau_0$.
This means that we can regard $|\psi_0({\mathbf x},t^\prime)|^2$
as being equal to $|\psi_0({\mathbf x}, t)|^2$
in the expression (\ref{eq:gp}) of $\Gamma_P$, i.e
\begin{equation}
|\psi_0({\mathbf x},t^\prime)|^2 = |\psi_0({\mathbf x}, t)|^2 \equiv
p_0({\mathbf x},t) \,.
\end{equation}
We apply the Fourier transformations
\begin{eqnarray}
\tilde p_0 ({\mathbf k}) &\equiv& \int p_0 ({\mathbf x}) e^{-i{\mathbf k x}}
d^3 x \, , \nonumber\\
\tilde G({\mathbf k}) &\equiv& \int G({\mathbf x}) e^{-i{\mathbf k x}} d^3 x
\, ,
\end{eqnarray}
and use the convolution theorem to obtain
\begin{equation}
\gamma_P = \frac{1}{2(2\pi)^3} \int \limits_{-\tau_0}^{+\tau_0} \int
|\tilde p_0({\mathbf k},t)|^2 \tilde G({\mathbf k})\,
e^{i c \tau k_z}\, d^3k\,d\tau
\, .
\end{equation}
Because $c\tau_0$ is large compared with the extension of the
condensate we can approximate
\begin{equation}
\gamma_P=\frac{\omega_0^{-1}}{2 (2\pi)^3} \int
|\tilde p_0 ({\mathbf k},t) |^2
\tilde G({\mathbf k})\, \delta ({k_z}/{k_0})\, d^3 k \, .
\end{equation}
We utilize the definition (\ref{eq:g}) of the $G$ function and find that
\begin{equation}
\tilde G = \frac{\omega_0}{4} \frac{\chi_0^2}{\hbar c} I \tilde C({\mathbf
k}+k_0 {\mathbf e}_z)
\end{equation}
with $\tilde C ({\mathbf k})$ being the Fourier--transformed commutator
(\ref{eq:c}) expressed in terms (\ref{eq:cm}) of the mode density
$d\mu({\mathbf k})$. We obtain
\begin{equation}
\label{eq:gpmu}
\gamma_P=\frac{\pi}{4} \frac{\chi_0^2}{\hbar c} I \frac{1}{(2 \pi)^3} \int
|\tilde p_0 ({\mathbf k}-k_0{\mathbf e}_z,t)|^2
\delta(\frac{k_z}{k_0}-1)\,
d \mu({\mathbf k}) \, .
\end{equation}
Since a typical Bose--Einstein condensate extends over many optical
wave lengths, the Fourier--transformed spatial distribution $\tilde
p_0({\mathbf k}) $ is narrow compared with the optical mode density.
Furthermore, in Eq.\ (\ref{eq:gpmu}), $\tilde p_0({\mathbf k})$ is
moved by $k_0 {\mathbf e}_z$ to an area in ${\mathbf k}$ space where
it overlaps with the optical field. The spatial distribution of the
condensate thus probes the paraxial optical mode density. In this
case, see Section IID, the optical--field commutator acts as a
three--dimensional delta function.  We set $d\mu ({\mathbf k})=d^3
k$ and obtain, finally,
\begin{equation}
\label{eq:gp1}
\gamma_P = \frac{\pi}{4} \frac{\chi_0^2}{\hbar c} I \frac{k_0}{(2 \pi)^3}
\int |\tilde p_0 (k_x,k_y,0,t)|^2\, dk_x \,dk_y \, ,
\end{equation}
or
\begin{equation}
\label{eq:gp2}
\gamma_P = \frac{\pi}{4} \frac{\chi_0^2}{\hbar c} I \lambda^{-1} \int \left(
\int \limits_{-\infty}^{+\infty} p_0({\mathbf x},t) dz \right)^2 dx\,dy \, .
\end{equation}
In this way we have expressed the phase--diffusion rate
$\gamma_P$ in terms of the spatial probability distribution
$p_0({\mathbf x},t)$ for a single condensed atom.

\subsection{Depletion}

The second backaction process of phase--contrast imaging is a familiar
bosonic depletion. The Liouvillian ${\cal L}_2 \hat \rho_0$ contains a
Hamiltonian phase shifting that is proportional to $\mbox{Im}
\Gamma_L-\mbox{Im} \Gamma_P$ and a depletion that occurs at the rate
$\gamma_L-\gamma_P$ with $\gamma_L\equiv \mbox{Re} \Gamma_L$.
We have calculated $\gamma_P$ in the previous section, let us turn to
$\gamma_L$ here. First, we proceed along similar lines as in the calculation
of $\gamma_P$ and get
\begin{equation}
\label{eq:gl1}
\gamma_L = \frac{\pi}{4} \frac{\chi_0^2}{\hbar c} I \frac{1}{(2 \pi)^3} \int
\delta({k_z}/{k_0}-1)\, d \mu({\mathbf k}) \, .
\end{equation}
Because
\begin{equation}
|\tilde p_0({\mathbf k})|=
\left| \int p_0({\mathbf x})e^{i {\mathbf k x}} d^3 x
\right| \leq \int p_0({\mathbf x}) d^3 x = 1
\end{equation}
we see from Eqs. (\ref{eq:gpmu}) and (\ref{eq:gl1}) that
\begin{equation}
\gamma_L \geq \gamma_P \,.
\end{equation}
The overall depletion rate $\gamma_L-\gamma_P$ is non--negative.
This satisfactory result illustrates the consistency of our theory.

Note, that we cannot replace the mode density $d\mu({\mathbf k})$ in Eq.
(\ref{eq:gl1}) by $d^3 k$, as we have done in the case of phase diffusion,
because in Eq.\ (\ref{eq:gl1}) we integrate $d\mu({\mathbf k})$ over a broad
spectral range. Therefore, we must use the broad--spectrum behavior
(\ref{eq:cb}) of the commutator (\ref{eq:cm}) to calculate $\gamma_L$. We
insert the definition (\ref{eq:g}) of $G$ into the expression (\ref{eq:gl})
of $\Gamma_L$ and obtain for $\gamma_L= \mbox{Re} \Gamma_L$ the integral
\begin{equation}
\label{eq:gl2}
\gamma_L= \frac{k_0}{8} \frac{\chi_0^2}{\hbar c} I
\int \limits_{-c \tau_0}^{+c\tau_0} C(z {\mathbf e}_z) e^{-ik_0 z} d z \, .
\end{equation}
Then we apply the explicit formula (\ref{eq:cb}) of the commutator and get
\begin{eqnarray}
\gamma_L
&=&
\frac{\pi}{4 i} \frac{\chi_0^2}{\hbar c} I \lambda^{-3} \int
\limits_{-c\tau_0}^{+c \tau_0} \frac{d z}{z-i 0} \nonumber\\
&=& \frac{\pi^2}{2} \frac{\chi_0^2}{\hbar c} I \lambda^{-3} \frac{1}{2 \pi i}
\left( \oint \frac{dz}{z-i0} - \int_{(2)} \frac{dz}{z} \right) \, ,
\label{eq:contours}
\end{eqnarray}
with the contours indicated in Fig.\ 2.
We use the residue theorem for the first integral and calculate the second
explicitly. We obtain, finally,
\begin{equation}
\label{eq:gl3}
\gamma_L= \frac{\pi^2}{4} \frac{\chi_0^2}{\hbar c} I \lambda^{-3} \, .
\end{equation}
Our result is independent on the value of the integration time $\tau_0$ and is
therefore consistent with the assumed Markovian behavior of the quantum
backaction.
Furthermore, Eq.\ (\ref{eq:gl3}) does not depend on the spatial shape
of the condensate.
The loss rate is universal.

Equation (\ref{eq:gl2}) shows that the depletion rate $\gamma_L$
depends on the optical--field commutator $C(x,y,z)$ at $x,y=0$
integrated over $z$. So, if the commutator were a three--dimensional
delta function, as in usual quantum optics, our calculations would
give an infinite backaction, a result in clear contradiction to
experimental evidence \cite{Andrews}. At this point the reader will
appreciate the reason why we have developed a quantum theory of
paraxial propagation in Section II. This theory employs both a
biquadratic interaction (\ref{eq:hi}) of light and matter and a
modified commutator (\ref{eq:cb}) that produces a finite backaction.

On the other hand, a first--order perturbation theory (\ref{eq:hlow})
of canonical electromagnetism (\ref{eq:h}) is also biquadratic in the
light--matter interaction. Here the canonically conjugate variables are
${\mathbf A}$ and ${\mathbf D}$ and hence we should postulate that the
commutator of $\mathbf{\hat A}$ and $\mathbf{\hat D}$ is proportional
to a transversal delta function.
However, this approach produces an infinite backaction. Why is
perturbation theory not appropriate in the canonical Hamiltonian
(\ref{eq:h})?
Remember that here we approximate the light--matter interaction
$(1+\chi_0 |\psi|^2)^{-1}$ by the first two terms of the geometric
series $\sum_{\nu=0}^\infty (-\chi_0|\psi|^2)^\nu$.
If $\psi$ is an operator $\hat \psi$, as it should be in the quantum
statistics of matter waves, the particle nature of matter will make the
density operator $\hat \psi^\dagger \hat \psi$ potentially large,
regardless how small $\chi_0$ is. (For example, the density of a
localized particle is a three-dimensional delta function.)
Therefore, we must not approximate $(1+\chi_0\hat \psi^\dagger \hat
\psi )^{-1}$ by $1-\chi_0 \hat \psi^\dagger \hat \psi$ and, moreover,
we must not see $(1+\chi_0\hat \psi^\dagger \hat \psi)^{-1}$  as a
geometric series at all.

Let us add a more intuitive picture to our mathematical
arguments. Phase--contrast imaging involves a particle--particle
scattering of atoms and photons that depends on the ability of the
incident photons to localize. The optical--field commutator describes
potential photon fluctuations and therefore photon localization as
well \cite{MWL}. Mandel and Wolf \cite{MWL} discuss in great detail
that photons can manifest themselves as particles only within a volume
that is bounded by the optical wave length.
The finite localization volume of photons thus leads to a finite quantum
backaction.

\subsection{Estimation of the effect}

How large is the backaction?
Let us estimate solely the order of magnitude. First, we compare the
depletion rate $\gamma_L$ with the phase--diffusion rate
$\gamma_P$. The latter depends on the spatial shape of the
condensate. To find a rough estimate for $\gamma_P$ we assume that the
condensate extends over $a_x$, $a_y$, and $a_z$ in $x$, $y$, and $z$
direction, respectively, and we model the Fourier--transformed
probability distribution by the Gaussian
\begin{equation}
  \label{eq:gauss}
  \tilde p_0 ({\mathbf k}) =
  \exp\left( - {\textstyle\frac{1}{2}} a_x^2 k_x^2 -
  {\textstyle\frac{1}{2}} a _y^2 k_y^2 -
  {\textstyle\frac{1}{2}} a_z^2 k_z^2 \right) \, .
\end{equation}
We obtain from Eq.\ (\ref{eq:gp1}) the phase--diffusion rate
\begin{equation}
  \gamma_P = \frac{\pi^2}{4} \frac{\chi_0^2}{\hbar c} I \frac{1}{
  \lambda (2 \pi a_x) (2 \pi a_y)} \, ,
\end{equation}
or, expressed in terms of the depletion rate $\gamma_L$ via
Eq.\ (\ref{eq:gl3}),
\begin{equation}
  \gamma_P = \gamma_L \frac{\lambda^2}{(2 \pi a_x)(2 \pi a_y)} \, .
\end{equation}
A typical Bose--Einstein condensate extends over many optical wave
lengths $\lambda$.
Therefore,
\begin{equation}
  \gamma_L \gg \gamma_P \, .
\end{equation}
Depletion far outweighs phase diffusion in the backaction of
phase--contrast imaging.

The depletion rate (\ref{eq:gl3}) is proportional to the intensity of
the illumination. A moderate intensity will thus lead to a low
backaction.
On the other hand, low intensity light exhibits large quantum--optical
phase fluctuations. To give a rough estimation, we use the uncertainty
product \cite{criticalremark}
\begin{equation}
\label{eq:uncertain}
  \delta \phi \, \delta n \approx 1
\end{equation}
of the phase and photon--number fluctuations. A coherent--state
illumination has Poissonian photon statistics with a variance
$\delta^2 n$ that equals the mean $\bar n$, i.e.
\begin{equation}
  \label{eq:noise}
  \delta^2\phi \approx {\bar n}^{-1} \, .
\end{equation}
The mean photon number is the product of the average photon density
$I/(\hbar \omega_0 c)$ with the optical mode volume
during the observation time $\Delta t$.
The mode volume is roughly given by a cylinder of radius $\lambda$
and length $c\,\Delta t$.
Therefore, we obtain
\begin{equation}
  \bar n \approx
  \frac{I}{\hbar\omega_0c}\,\pi\lambda^2c\,\Delta t =
  \pi\,\frac{\lambda^2I\Delta t}{\hbar\omega_0}\, .
\end{equation}
One can see an image of a phase object only when the produced phase
shift $\Delta\phi$ exceeds the phase noise $\delta \phi$.
Let us estimate the signal $\Delta\phi$ for a condensate of $N$
atoms with single--atom probability distribution $p_0({\mathbf x})$:
\begin{eqnarray}
  \label{eq:signal}
  \Delta \phi &=& \int ({\mathbf k}- {\mathbf k}_0) d{\mathbf x}
  \nonumber\\
  &\approx& \frac{2 \pi}{\lambda} \int \left( \sqrt{1+\chi_0 N
  p_0({\mathbf x})} -1 \right) dz \nonumber\\
  &\approx& \frac{\pi}{\lambda} \chi_0 N \eta
\end{eqnarray}
where $\eta$ denotes the effective two--dimensional density
\begin{equation}
  \eta = \int p_0 ({\mathbf x}) dz \, .
\end{equation}
For our simple model (\ref{eq:gauss}) the largest value of $\eta$ is
\begin{equation}
  \eta = \frac{1}{2 \pi a_x a_y} \, .
\end{equation}
Let us compare the signal--to--noise
ratio $\Delta \phi / \delta \phi$ with the induced backaction.
According to our theory, the incident light depletes gradually
the condensate during the observation time $\Delta t$.
The average number of atoms,
$\mbox{tr}\{\hat a_0^\dagger \hat a_0\,\hat \rho_0\}$,
decays by the factor of $\exp(-2\gamma_L\Delta t)$.
We express our result (\ref{eq:gl3}) for the backaction rate
$\gamma_L$ in terms of the signal--to--noise
ratio $\Delta \phi / \delta \phi$ using the estimations
(\ref{eq:noise}-\ref{eq:signal}), and find
\begin{equation}
\label{eq:kappa}
  \kappa \equiv 2\gamma_L \Delta t \approx
  \left( \frac{\Delta \phi}{\delta \phi}\right)^2
  \left(N \lambda^2 \eta\right)^{-2} \, .
\end{equation}
The more atoms are condensed the lower is the required intensity for
producing a faithful image and, consequently, the lower is the induced
backaction. On the other hand, the more macroscopic the wave function
is the lower is the probability density $\eta$ and the larger is the
backaction rate.
We also realize from Eq.\ (\ref{eq:kappa}) that the depletion 
factor $\exp(-\kappa)$ is independent of the susceptibility $\chi_0$
(and in particular of the detuning from atomic resonances),
given a fixed signal--to--noise ratio $\Delta \phi / \delta \phi$.
This is easy to understand, because
for a large $\chi_0$ the signal $\Delta \phi$ is large, 
but so is the backaction (\ref{eq:gl3}).

We obtain for the critical value $\Delta \phi \approx \delta \phi$
the depletion constant $\kappa \approx (N\lambda^2\eta)^{-2}$.
Let us assume that $a_x \, a_y$ is roughly $10^4\lambda^2$.
In this case the depletion constant $\kappa$ is about $N^{-2} 10^{10}$.
A small condensate \cite{Hulet} with $N\approx 10^3$ atoms disappears
immediately! Therefore, a continuous monitoring of the condensate
is impossible, in agreement with experimental observation
\cite{Hulet}. Larger condensates are more robust.
For $N\approx10^6$ atoms \cite{Andrews} we obtain a
decay factor $\exp(-\kappa)$ of about $1-\kappa \approx 1 - 10^{-2}$.
The depletion is in the order of one percent
which agrees well with the experimental observation \cite{Andrews}.
However, the authors of Ref.\ \cite{Andrews}
have interpreted the decay of the condensate
as an residual absorption effect.
Our calculations seem to indicate that the quantum backaction
of phase--contrast imaging 
has caused the observed depletion
via the momentum transfer of the illuminating light quanta.

Finally, we remark that the use of appropriately squeezed
\cite{squeezing} illumination would enhance the signal--to--noise ratio
at a given, moderate quantum backaction.

\section{Summary}

Phase--contrast imaging \cite{Andrews} induces a quantum backaction
on Bose--Einstein condensates.
The backaction consists of two dissipative processes:
a phase diffusion and a gradual depletion of the condensate.
If phase--contrast imaging were a quantum nondemolition measurement
of the condensate density, solely phase diffusion would occur.
We have shown, however, that the depletion is the dominant process.
Our estimations indicate that the quantum backaction is indeed
an observable effect that limits the otherwise destructionless
character of phase--contrast imaging \cite{Andrews}.
Condensates with a few number of atoms are especially vulnerable
to optical observations.

In our theoretical study we ignored entirely the absorption
of light (the imaginary part of the refractive index)
and we focused our attention on the quantum backaction.
This is justified in the limit of strong detuning,
because here the absorption virtually vanishes.
Note that our theory is only valid as long as the condensate
is still significantly populated.
We have assumed {\it a priori} the existence 
and the dominance of the condensate.

We developed a quantum theory of paraxial light propagation
in many--atom samples that are off--resonant with respect
to the light.
This theory goes beyond the immediate purpose of the present paper
and may find wider application in the fascinating subject of
combining concepts of quantum measurement theory with
quantum--optical propagation and the statistical theory
of quantum gases.

\section*{Acknowledgments}

We acknowledge the support of the research consortium
{\it Quantum Gases} of the
Deutsche Forschungsgemeinschaft.
U.\ L.\ thanks the Alexander von Humboldt Foundation and the
G\"oran Gustafsson Stiftelse.
T.\ K.\ was supported by an E\"otv\"os Fellowship and by the grants
F017381 and T023777 of the National Research Fund of Hungary
(OTKA).
T.\ K.\ is grateful to W.\ P.\ Schleich for his kind hospitality in Ulm.

\section*{Appendix A}

In this appendix we calculate explicitly the optical field commutator
$C({\mathbf x})$ and we discuss a few general properties of
$C({\mathbf x})$.
We start from the expression (\ref{eq:cm}) of $C({\mathbf x})$ in terms
of the mode density $d\mu({\mathbf k})$,
\begin{eqnarray}
C({\mathbf x}) &=& \frac{1}{(2 \pi)^3}
\int e^{i {\mathbf k x}} d\mu({\mathbf k}) \,,
\nonumber\\
d\mu({\mathbf k}) &=&
\delta \left( k_z/k_0 +
{\textstyle\frac{1}{2}} (k_x^2+k_y^2) / k_0^2-1 \right)\,
d^3 k \, .
\label{eq:cadef}
\end{eqnarray}
To calculate the Fourier integral we introduce dimensionless
variables, $k_x=k_0\,\zeta_x$, $k_y=k_0\,\zeta_y$ with
\begin{equation}
k_z = k_0 - {\textstyle\frac{1}{2}}k_0(\zeta_x^2+\zeta_y^2)
\, ,
\end{equation}
and obtain
\begin{eqnarray}
C({\mathbf x}) = \frac{k_0^3}{(2\pi)^3} & e^{ik_0 z} &
\int_{-\infty}^{+\infty}
e^{-i\zeta_x^2 k_0 (z-i0)/2 + i k_0 x \zeta_x}\, d\zeta_x
\nonumber\\
&\times&
\int_{-\infty}^{+\infty}
e^{-i\zeta_y^2 k_0 (z-i0)/2 + i k_0 y \zeta_y}\, d\zeta_y
\,.
\label{eq:integrals}
\end{eqnarray}
In Eq.\ (\ref{eq:integrals}) we have given $z$ an
infinitesimally small yet negative imaginary part $-i0$
to ensure that the integrals converge.
The calculation of Eq.\ (\ref{eq:integrals}) is straightforward,
and we obtain the result
\begin{equation}
\label{eq:cexplicit}
C({\mathbf x}) =
\frac{k_0^2}{i(2 \pi)^2(z-i 0)}
\exp \left[\frac{i k_0}{2 z}(x^2 + y^2)+ik_0 z \right] \,.
\end{equation}
Let us discuss a few elementary properties of the paraxial
commutator.
First, we see from the mode density in Eq.\ (\ref{eq:cadef})
that $C({\mathbf x})$ obeys the partial differential equation
\begin{equation}
\left( -i \frac{\partial}{\partial z} - \frac{\nabla^2_\perp}{2 k_0} \right)
C =  k_0 C \,.
\end{equation}
The commutator behaves like a paraxial monochromatic wave
that travels freely in space without being influenced by a medium.
Furthermore, when $z$ approaches zero, the commutator
$C(x,y,z\rightarrow 0)$ vanishes effectively for $x,y\neq 0$,
because the exponential function in formula (\ref{eq:cexplicit})
is highly oscillating in this limit.
To see what happens at $x,y=0$ we integrate
$C(x,y,z\rightarrow 0)$ with respect to $x$ and $y$.
Since
\begin{eqnarray}
\int_{-\infty}^{+\infty} \int_{-\infty}^{+\infty} &&
\exp \left[\frac{i k_0}{2 z}(x^2 + y^2)\right]
dx\,dy
\nonumber\\
&=& 2\pi \int_0^\infty
\exp \left(\frac{i k_0}{2 z}r^2\right) d({\textstyle\frac{1}{2}}r^2)
\nonumber\\
&=& 2\pi i \frac{z}{k_0}
\end{eqnarray}
we obtain, finally,
\begin{equation}
C(x,y,z \to 0) = \frac{k_0}{2 \pi} \, \delta(x)\,\delta(y) \, .
\end{equation}
The commutator $C({\mathbf x})$ appears as a paraxial Green's
function.

\section*{Appendix B}
In this appendix we derive the master equation (\ref{eq:master0})
that describes the quantum backaction of phase--contrast imaging
on Bose--Einstein condensates.
We start from the general master equation (\ref{eq:master}) and
proceed in two steps.
First, we reformulate the general master equation (\ref{eq:master})
utilizing the fact (\ref{eq:pw}) that the incident light
is a plane wave 
and using the conservation (\ref{eq:ac}) of the total
number of atoms.
Then we average the resulting Liouvillian with respect to the
above--condensate part. This, finally, gives us the desired
master equation for the condensate.

To begin, let us first consider the Hamiltonian term
$\mbox{tr}_R [\hat \rho_I \hat \rho_{IR} , \hat H_{II}]$
in Eq.\ (\ref{eq:master})
that describes the average light force on the atomic sample.
We calculate the commutator of the interaction Hamiltonian
and the total density matrix in the Schr\"odinger picture
using Eq.\ (\ref {eq:pw})
and the conservation of the total number of atoms, Eq.\ (\ref{eq:ac}),
\begin{equation}
\mbox{tr}_R[\hat H_I,\hat\rho \hat\rho_R]=\frac{I}{2 c}\chi_0\,
[\int \hat\psi^{\dagger}\hat\psi\, d^3 x, \hat \rho\,]=0
\,\,.
\end{equation}
Consequently, we get in the interaction picture
\begin{equation}
\mbox{tr}_R[\hat H_{II},\hat\rho_I \hat\rho_{IR}]=0 \,\,.
\end{equation}
The overall light force of an uniform illumination vanishes.

Let us turn to the Liouvillian part ${\cal L} \hat \rho_I$
of the master equation (\ref{eq:master}).
First we calculate the correlation functions (\ref{eq:pw})
of the incident light in the interaction picture.
We neglect optical diffraction inside the atomic sample,
see Eq.\ (\ref{eq:nodiff}), and apply Eq.\ (\ref{eq:pw})
to get
\begin{eqnarray}
&\mbox{tr}_R&\left\{
\hat  \varphi_I^{\dagger} ({\bf x_1},t_1)
\hat \varphi_I({\bf x_2},t_2) \, \hat
\rho_{IR}(t_2) \right\}
\nonumber\\
&=&
\mbox{tr} _R \left\{
\hat  \varphi_I^{\dagger} ({\bf x_1}-c(t_1-t_2){\bf e}_z,t_2)
\hat \varphi_I({\bf x_2},t_2) \, \hat
\rho_{IR}(t_2) \right\}
\nonumber
\\
& = &
\mbox{tr} _R \left\{
e^{-i \hat H_R t_2}
\hat  \varphi^{\dagger} ({\bf x_1}-c(t_1-t_2){\bf e}_z)
\hat \varphi({\bf x_2}) \, \hat
\rho_R(t_2)\,e^{i \hat H_R t_2} \right\}
\nonumber
\\
& = &
\mbox{tr} _R \left\{
\hat  \varphi^{\dagger} ({\bf x_1}-c(t_1-t_2){\bf e}_z)
\hat \varphi({\bf x_2}) \, \hat
\rho_R(t_2) \right\}
\nonumber
\\
& = &
{I \over \hbar \omega_0 c}\,
\exp[{i k_0 (z_2-z_1+c(t_1-t_2))}]
\,\,,
\label{eq:cc1}
\end{eqnarray}
and along similar lines
\begin{eqnarray}
& \mbox{tr}_R&
\left\{
\hat \varphi_I^{\dagger}({\bf x_1},t_1)
\hat \varphi_I^{\dagger} ({\bf x_2},t_2)
\hat \varphi_I ({\bf x_1},t_1)
\hat \varphi_I ({\bf x_2},t_2)\,
\hat \rho_{IR} \right\}
\nonumber\\
& = & \mbox{const}
\,\, .
\label{eq:cc2}
\end{eqnarray}
To evaluate the Liouvillian ${\cal L} \hat \rho_I$ in Eq.\ (\ref{eq:master})
we calculate the double commutator
using the shorthand notation
$\hat\psi_i\equiv\hat\psi_I({\bf x}_i,t_i)$,
$\hat\varphi_i\equiv\hat\varphi_I({\bf x}_i,t_i)$,
\begin{eqnarray}
\label{eq:dc}
\lefteqn{\mbox{tr}_R\left\{
[\hat H_{II}(t_1),[\hat H_I(t_2),\hat\rho_I \hat\rho_{IR}]]\right\}}
\nonumber\\
&=&\frac{\hbar^2 \omega_0^2}{4} \chi_0^2 \int \int \mbox{tr}_R\left\{
\hat\psi_1^{\dagger}\hat\psi_1\,\hat\varphi_1^{\dagger}\hat\varphi_1
\,
\hat\psi_2^{\dagger}\hat\psi_2\,\hat\varphi_2^{\dagger}\hat\varphi_2
\,
\hat\rho_I\hat\rho_{IR}
\right.
\nonumber\\
&&- \,
\hat\psi_1^{\dagger}\hat\psi_1\,\hat\varphi_1^{\dagger}\hat\varphi_
1\,
\hat\rho_I\hat\rho_{IR}\,
\hat\psi_2^{\dagger}\hat\psi_2\,\hat\varphi_2^{\dagger}\hat\varphi_2
\nonumber\\
&&- \,\hat\psi_2^{\dagger}\hat\psi_2\,
\hat\varphi_2^{\dagger}\hat\varphi_2\,
\hat\rho_I\hat\rho_{IR}\,
\hat\psi_1^{\dagger}\hat\psi_1\,\hat\varphi_1^{\dagger}\hat\varphi_1
\nonumber\\
&&\left.
+ \,\hat\rho_I\hat\rho_{IR}\,
\hat\psi_2^{\dagger}\hat\psi_2\,\hat\varphi_2^{\dagger}\hat\varphi_2
\,
\hat\psi_1^{\dagger}\hat\psi_1\,\hat\varphi_1^{\dagger}
\hat\varphi_1\right\}
\nonumber\\
&&\times\,
d^3 x_1\, d^3 x_2\nonumber\\
&=&\frac{\hbar^2 \omega_0^2}{4} \chi_0^2 \int \int
\left(\hat\psi_1^{\dagger}
\hat\psi_1\,\hat\psi_2^{\dagger}\hat\psi_2\,\hat\rho_I-
\hat\psi_2^{\dagger}\hat\psi_2\,\hat\rho_I\,\hat\psi_1^{\dagger}
\hat\psi_1\right)
\nonumber\\
&& \times\,
\mbox{tr}_R\left\{\hat\varphi_1^{\dagger}\hat\varphi_1\,
\hat\varphi_2^{\dagger}\hat\varphi_2\,
\hat\rho_{IR}\right\}d^3 x_1d^3 x_2  + \mbox{H.c.}\,\,.
\end{eqnarray}
Then we utilize the commutator relation of the radiation
field (\ref{eq:c}) and the diffraction--less propagation
(\ref{eq:nodiff}) in the interaction zone to write
$\mbox{tr}_R\{\hat\varphi^{\dagger}_1\hat\varphi_1\,
\hat\varphi^{\dagger}_2\hat\varphi_2\,\hat\rho_{IR}\}$
in normal order,
\begin{eqnarray}
&\mbox{tr}_R&\left\{\hat\varphi^{\dagger}_1\hat\varphi_1\,
\hat\varphi^{\dagger}_2\hat\varphi_2\,\hat\rho_{IR}\right\}
\nonumber\\
&=&
\mbox{tr}_R\left\{\hat\varphi^{\dagger}_1\,\hat\varphi^{\dagger}_2
\hat\varphi_1\hat\varphi_2\,\hat\rho_{IR}\right\}
\nonumber\\
&& + \,
\mbox{tr}_R\left\{\hat\varphi^{\dagger}_1\hat\varphi_2\,
\hat\rho_{IR}\right\}C({\bf x_1}-{\bf x_2}-c(t_1-t_2){\bf e}_z)
\,\,.
\end{eqnarray}
The normally--ordered correlation function (\ref{eq:cc2})
is constant, and so 
$\mbox{tr}_R\{\hat\varphi^{\dagger}_1\hat\varphi^{\dagger}_2
\hat\varphi_1\hat\varphi_2\hat\rho_{IR}\}$
does not contribute to
the double commutator (\ref {eq:dc}),
when we take into account the atom--number conservation (\ref{eq:ac}).
To proceed with the remaining term, we utilize Eq.\ (\ref{eq:cc1})
to express the correlation function
$\mbox{tr}_R\{\hat\varphi^{\dagger}_1\hat\varphi_2\,
\hat\rho_{IR}\}$,
define in Eq.\ (\ref{eq:g}) the function $G({\bf x})$,
and get
\begin{eqnarray}
{\cal L}\hat\rho_I
&=&
\int_{-\infty}^t\int\int \left(\hat\psi_1^{\dagger}
\hat\psi_1\hat\psi_2^{\dagger}\hat\psi_2\hat\rho_I-
\hat\psi_2^{\dagger}\hat\psi_2\hat\rho_I\hat\psi_1^{\dagger}
\hat\psi_1\right)
\nonumber\\
&& \times\,
G({\bf x_1}-{\bf x_2}
-c(t-t^{\prime}){\bf e}_z) \,d^3 x_1\, d^3 x_2\, dt'
+ \mbox{H.c.} \,\,.
\label{eq:liou}
\end{eqnarray}
In this way we have found the Liouvillian that describes the quantum
backaction of an uniform illumination on the total atomic sample.

So far we have not made any essential
assumption on the state of the atoms and in particular we have not
exploited the fact that the atoms are Bose--condensed. To incorporate
Bose--Einstein condensation we expand the atomic annihilation operator
$\hat\psi_I$ in the interaction picture in a complete, orthonormal basis
of atomic modes, see Eq.\ (\ref{eq:atoms}).
We assume that only $\psi_0$ is significantly occupied and average
with respect to the above--condensate part being in the vacuum state.
(We utilize that $\hat a_\nu \hat \rho_I$ vanishes for $ \nu \neq 0$.)
For this we must
calculate some atomic correlation functions
that occur in the Liouvillian (\ref{eq:liou}).
We abbreviate $\mbox{tr}_{AC}\hat\rho_I$ by $\hat\rho_0$, and find
\begin{eqnarray}
&\mbox{tr}_{AC}&
\left\{\hat\psi_1^{\dagger}\hat\psi_2^{\dagger}\hat\psi_1
\hat\psi_2\,\hat\rho_I\right\}
\nonumber\\
&=&
(\hat a_0^{\dagger})^2\hat a_0^2 \,
|\psi_0({\bf x}_1,t_1)|^2\,|\psi_0({\bf x}_2,t_2)|^2\,\hat\rho_0
\,\,,
\label{eq:cfa}
\\
&\mbox{tr}_{AC}&\left\{\hat\psi_1^{\dagger}\hat\psi_1\hat\rho_I\right\}
=
\hat a_0^{\dagger}\hat a_0 |\psi_0({\bf x},t)|^2\hat\rho_0
\,\,,
\\
&\mbox{tr}_{AC}&\left\{\hat\psi_2^{\dagger}\hat\psi_2\,\hat\rho_I\,
\hat\psi_1^{\dagger}\hat\psi_1\right\}
\nonumber\\
&=&\mbox{tr}_{AC}\left\{\hat
\psi_2^{\dagger}\hat a_0 \,\hat\rho_I \,\hat
a_0^{\dagger}\hat\psi_1\right\}
\psi_0({\bf x}_2,t_2)\psi_0^{\ast}({\bf x}_1,t_1)
\,\,,
\\
&\mbox{tr}_{AC}&\left\{\hat\psi_2^{\dagger}\hat a_0 \,
\hat\rho_I\,\hat a_0^{\dagger}
\hat\psi_1\right\}
\nonumber\\
&=&
\sum_{\nu\mu} \mbox{tr}_{AC}\left\{\hat a_{\nu}^{\dagger}\hat
a_0\,\hat\rho_I
\,\hat a_0^{\dagger}\hat a_{\mu}\right\}
\psi_{\nu}^{\ast}({\bf x}_2,t_2)\psi_{\mu}({\bf x}_1,t_1)\nonumber\\
&=&\sum_{\nu}\mbox{tr}_{AC}\left\{\hat a_{\nu}^{\dagger}\hat
a_0\,\hat\rho_I\,
\hat a_0^{\dagger}\hat a_{\nu}\right\}
\psi_{\nu}^{\ast}({\bf x}_2,t_2)\psi_{\nu}({\bf x}_1,t_1)\nonumber\\
&=&\hat a_0^{\dagger} \hat a_0\, \hat\rho_0 \,\hat a_0^{\dagger}\hat a_0 \,
\psi_0^{\ast}
({\bf x}_2,t_2)\psi_0({\bf x}_1,t_1)
\nonumber\\
&&+\,
\hat a_0\hat\rho_0\hat a_0^{\dagger}
\sum_{{\nu}\not=0}\psi_{\nu}^{\ast}({\bf x}_2,t_2)\psi_{\nu}({\bf x}_1,t_1)
\nonumber\\
&=&\hat a_0^{\dagger}\hat a_0\,\hat\rho_0\,
\hat a_0^{\dagger}\hat a_0\,\psi_0^{\ast}
({\bf x}_2,t_2)\psi_0({\bf x}_1,t_1)+\hat a_0\hat\rho_0\hat a_0^{\dagger}\,
\delta^{(3)}({\bf x}_1
-{\bf x}_2)\nonumber\\
&&-\, \hat a_0\hat\rho_0\hat a_0^{\dagger}\,
\psi_0^{\ast}({\bf x}_2,t_2)
\psi_0({\bf x}_1,t_1)
\,\,.
\label{eq:cfz}
\end{eqnarray}
We apply the results (\ref{eq:cfa}-\ref{eq:cfz})
to obtain after averaging of the
Liouvillian (\ref{eq:liou}) with respect to the
above--condensate part the final master equation
(\ref{eq:master0}).

\newpage

\section*{Figure caption}
{\bf Figure 1:}
Scheme of phase--contrast imaging \cite{Hecht}.
To produce an image {\it I} of a transparent object {\it O}
the non--scattered light is phase--shifted (phase--contrast microscopy)
or blocked (dark--ground method \cite {Andrews})
in the focal plane {\it F} of the magnifying lens {\it L}.

{\bf Figure 2:}
Integration contours of Eq.\ (\ref{eq:contours}) in the complex
$z$ plane. The closed contour (1) \& (2) encircles a
$(z-i0)^{-1}$ singularity (dot).


\begin{thebibliography}{99}

\bibitem{pioneers}
M. H. Anderson {\it et al.},
Science {\bf 269}, 198 (1995);
C. C. Bradley {\it et al.},
Phys. Rev. Lett. {\bf 75}, 1687 (1995);
K. B. Davis {\it et al.},
{\it ibid.} {\bf 75}, 3969 (1995).

\bibitem{Andrews}
M. R. Andrews {\it et al.},
Science {\bf 273}, 84 (1996).

\bibitem{Hulet}
C. C. Bradley, C. A. Sackett, and R. G. Hulet,
Phys. Rev. Lett. {\bf 78}, 985 (1997);
C. A. Sackett {\it et al.},
Appl. Phys. B {\bf 65}, 433 (1997).

\bibitem{Hecht}
E. Hecht,
{\it Optics} (Addison--Wesley, Reading, 1989).

\bibitem{bec_optics}
O. Morice, Y. Castin, and J. Dalibard,
Phys. Rev. A {\bf 51}, 3896 (1995).
For the theory of light propagation
in Bose condensates see also
B. V. Svistunov and G. V. Shlyapnikov,
Sov. Phys. JETP {\bf 70}, 460 (1990);
{\it ibid.} {\bf 71}, 71, (1990);
H. D. Politzer,
Phys. Rev. A {\bf 43}, 6444 (1991);
M. Lewenstein and L. You,
Phys. Rev. Lett. {\bf 71}, 1339 (1993);
J. Javanainen,
{\it ibid.} {\bf 72}, 2375 (1994);
M. Lewenstein {\it et al.},
Phys. Rev. A {\bf 50}, 2207 (1994);
L. You {\it et al.},
{\it ibid.} {\bf 51}, 4712 (1995); {\bf 53}, 329 (1996);
J. Javanainen and J. Ruostekoski,
{\it ibid.} {\bf 52}, 3033 (1995);
R. Graham and D. F. Walls,
Phys. Rev. Lett. {\bf 76}, 1774 (1996);
A. Csord\'{a}s {\it et al.},
Phys. Rev. A {\bf 54}, R2543 (1996);
J. Ruostekoski and J. Javanainen,
{\it ibid.} {\bf 55}, 513 (1997);
J. Ruostekoski and D. F. Walls,
{\it ibid.} {\bf 56}, 2996 (1997).

\bibitem{Media}
L. Kn\"oll, W. Vogel, and D.--G. Welsch,
Phys. Rev. A {\bf 36}, 3803 (1987)
R. J. Glauber and M. Lewenstein,
{\it ibid.} {\bf 43}, 467 (1991);
L. Kn\"oll, W. Vogel, and D.--G. Welsch,
{\it ibid.} {\bf 43}, 543 (1991);
L. Kn\"oll and D.--G. Welsch,
Progr. Quant. Electron. {\bf 16}, 135 (1992).

\bibitem{Marte}
M. A. M. Marte and S. Stenholm,
Phys. Rev. A {\bf 56}, 2940 (1997).

\bibitem{Carmichael}
See e.g.
H. Carmichael,
{\it An Open Systems Approach to Quantum Optics}
(Springer, Berlin, 1993),
Chapter 1.2.

\bibitem{DGPS}
See e.g. 
F. Dalfovo, S. Giorgini, L. P. Pitaevskii, and S. Stringari,
Rev. Mod. Phys. (in press).

\bibitem{MWV}
See Ref.\ \cite{MandelWolf},
Chapter 12.3 and Chapter 12.11.1.

\bibitem{MWC}
Mandel and Wolf
obtain a commutation relation of the vectorial
photon absorption operator $\hat{\mathbf V}$ in Eq.\ (12.11-5)
of Ref.\ \cite{MandelWolf}.
Our operator $\hat\varphi$ is one polarization component of
$\hat{\mathbf V}$.
We specialize Eq.\ (12.11-5) to one component and
get the commutation relation
$$
C({\bf x}) = {1 \over L^3} \sum_{[{\bf k}]} e^{i{\bf k x}} \equiv
{1 \over (2 \pi)^3} \int e^{i{\bf k x}} d \mu({\bf k}) \, .
$$
for
$C({\mathbf x}_1-{\mathbf x}_2) \equiv
[ \hat \varphi({\mathbf x}_1), \hat \varphi ^\dagger({\mathbf x}_2]$.
Here $L$ denotes the usual quantization length that tends to infinity.
We have introduced the mode density $d \mu({\bf k})$
to replace the sum of modes by an integral.

\bibitem{QMT}
For the theory of continuous quantum measurements see
C. M. Caves and G. J. Milburn,
Phys. Rev. A {\bf 36}, 5543 (1987).
See also
A. Barchielli,
Phys. Rev. A {\bf 34}, 1642 (1986)
and
V. P. Belavkin,
J. Phys. A {\bf 22}, L1109 (1989)
and references cited therein.
For an application to BEC see
J. R. Corney and G. J. Milburn,
Phys. Rev. A {\bf 58}, 2399 (1998).

\bibitem{kinetics}
Our theoretical approach is related to the quantum kinetic theory
of weakly interacting Bose gases, see
C. W. Gardiner and P. Zoller,
Phys. Rev. A {\bf 55}, 2902 (1997);
C. W. Gardiner {\it et al.},
Phys. Rev. Lett. {\bf 79}, 1793 (1997);
D. Jaksch {\it et al.},
Phys. Rev. A {\bf 56}, 575 (1997);
C. W. Gardiner and P. Zoller,
{\it ibid.} {\bf 58}, 536 (1998);
D. Jaksch {\it et al.},
{\it ibid.} {\bf 58}, 1450 (1998);
C. W. Gardiner {\it et al.},
Phys. Rev. Lett. {\bf 81}, 5266 (1998).

\bibitem{fluctuations}
Analytic results on the fluctuations of the above--condensate part
have been obtained by
S. Stringari,
Phys. Rev. Lett. {\bf 77}, 2360 (1996);
M. Fliesser {\it et al.},
Phys. Rev. A {\bf 56}, R2533 (1997);
P. \"Ohberg {\it et al.},
{\it ibid.} {\bf 56}, R3346 (1997).
For numerical studies see e.g.
M. Edwards {\it et al.},
Phys. Rev. Lett. {\bf 77}, 1671 (1996);
K. G. Singh and D. S. Rokhsar,
{\it ibid.} {\bf 77}, 1676 (1996);
L. You, W. Houston, and M. Lewenstein,
Phys. Rev. A {\bf 55}, R1581 (1997).

\bibitem{Phase}
See e.g.
U. Leonhardt,
{\it Measuring the Quantum State of Light},
(Cambridge University Press,
Cambridge, 1997),
Chapter 6.3.

\bibitem{MWL}
See e.g. the discussion in Chapter 12.11.
of Mandel and Wolf \cite{MandelWolf}
and the references cited therein.

\bibitem{criticalremark}
The uncertainty relation (\ref{eq:uncertain})
between number and phase holds only in an
approximative sense for quasiclassical
states of light.
For a critical review of the quantum--optical
phase problem see
R. Lynch,
Phys. Rep. {\bf 256}, 367 (1995).

\bibitem{squeezing}
See e.g.
G. Breitenbach {\it et al.},
Nature {\bf 387}, 471 (1997).

\bibitem{MandelWolf}
L. Mandel and E. Wolf,
{\it Optical coherence and quantum optics},
(Cambridge University Press,
Cambridge, 1995).

\end{thebibliography}
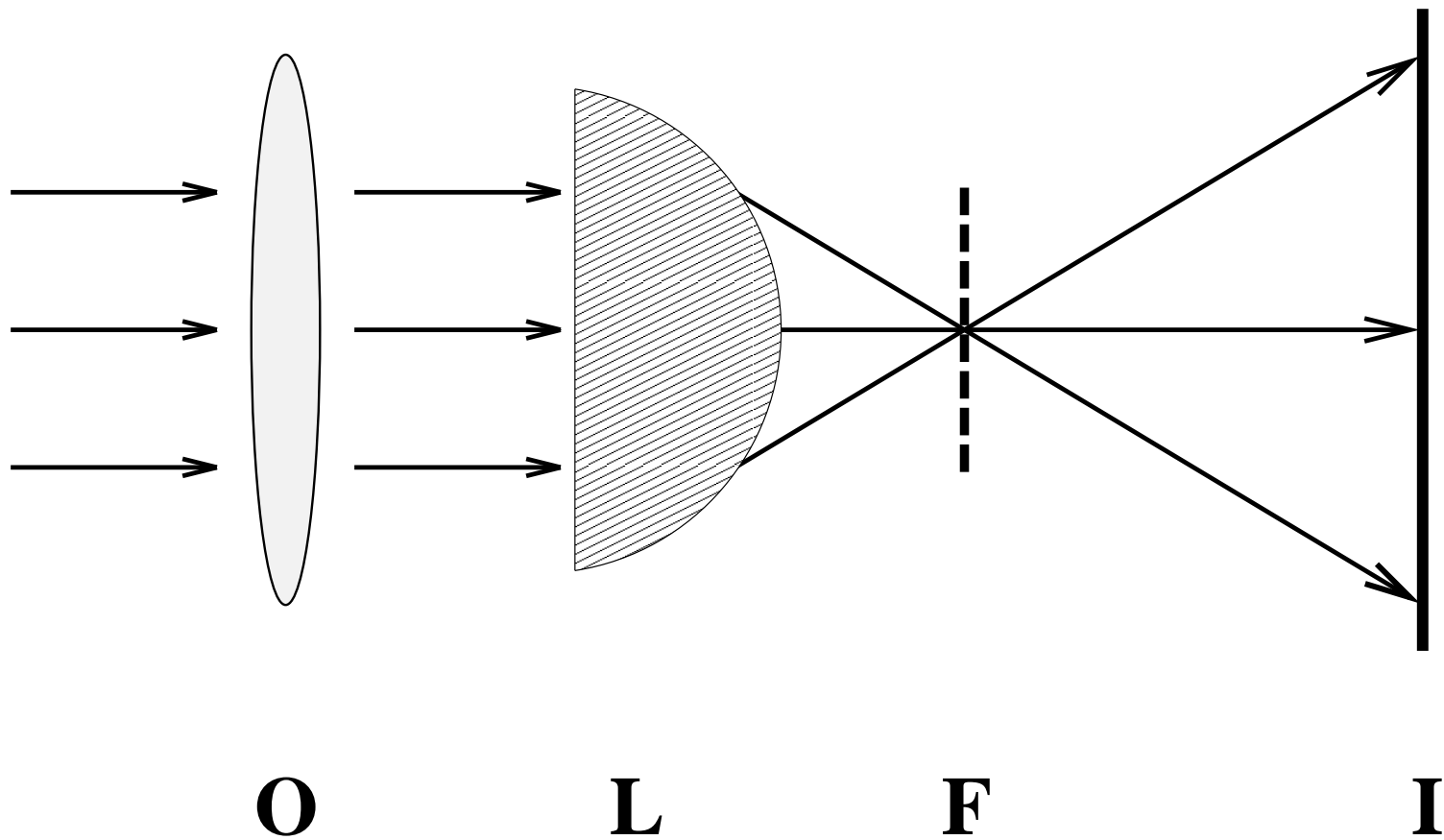
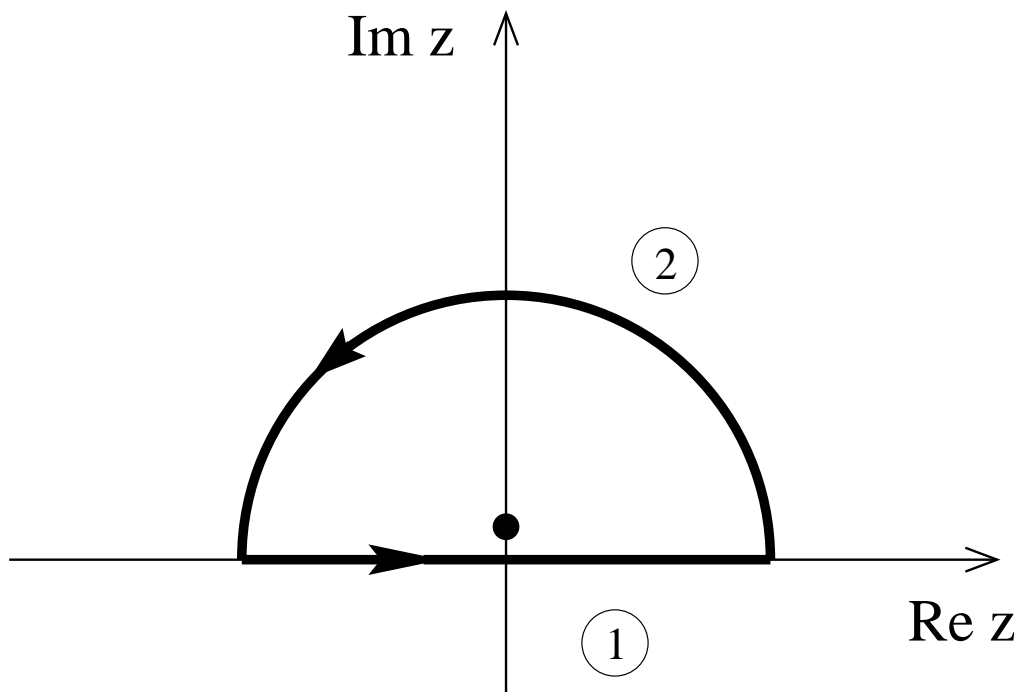
\end{document}